\documentclass[11pt]{article}
\usepackage{graphicx,natbib}
\bibpunct{(}{)}{;}{a}{}{;}

\newcommand{\D}{\hphantom{9}}
\newcommand{\DD}{\hphantom{99}}
\newcommand{\DDD}{\hphantom{999}}

\newcommand{\w}[1]{\hphantom{#1}}
\newcommand{\E}[1]{\mbox{$\times 10^{#1}$}}

\newcommand{\NOPRINT}[1]{\null}

\begin{document}
\title{Historical Reflections on the Work of\\ 
         IAU Commission 4 (Ephemerides)\\
         {\small To be Published in the Transactions of the International Astronomical Union,
         Volume XXIX-A}} 
\author{George H. Kaplan, editor\\[0.5ex]
     {\small Contributors: J.~A.~Bangert, A.~Fienga, W.~Folkner, C.~Hohenkerk, G.~H.~Kaplan, M.~Lukashova,}\\
     {\small E.~V.~Pitjeva, P.~K.~Seidelmann, M.~Sveshnikov, S.~Urban, J.~Vondr\'{a}k, J.~Weratschnig,
     J.~G.~Williams}}
\date{October 30, 2015}
\maketitle

\begin{abstract}
\noindent As part of a reorganization of the International Astronomical Union (IAU), Commission~4 (Ephemerides)
went out of existence after the IAU General Assembly in August 2015.  This paper presents brief discussions of
some of the developments in fundamental astronomy that have influenced and been influenced by the work of 
Commission~4 over its 96-year history.  The paper also presents notes about some of the publications of the
national institutions that have played an essential role in the commission's mission.   The contents of this paper were submitted for Commission~4's final report, to appear in IAU Transactions Vol.~XXIX-A.
\end{abstract}

\noindent Commission 4 was among the first set of commissions formed within the IAU at its founding in 1919.
(Commissions were originally called ``Standing Committees.'')  During its 96 years of service to the IAU and
astronomical community in general, the commission has been fortunate to have been
led by many distinguished scientists --- see the list of presidents below.

\smallskip
\begin{center}
\begin{tabular}{l | l | l}
\hline
\multicolumn{3}{c}{\bf Presidents of Commission 4} \\
\hline
1919--1922 \, P. H. Cowell             &  1967--1970  \, G. A. Wilkins        & 1991--1994 \,  B. D. Yallop  \\           
1922--1928 \, W. S. Eichelberger  &  1970--1973  \, J. Kovalevsky      & 1994--1997 \, H. Kinoshita \\                 
1928--1932  \, E. W. Brown              & 1973--1976  \, R. L. Duncombe & 1997--2000 \, E. M. Standish \\              
1932--1938  \, L. J. Comrie              & 1976--1979  \, V. K. Abalakin      &  2000--2003 \, J. Chapront \\                
1938--1952  \, G. Fayet                     & 1979--1982  \, A. M. Sinzi            & 2003--2006  \, G. Krasinsky \\          
1952--1958  \, D. H. Sadler              & 1982--1985  \, T. Lederle             & 2006--2009 \, T. Fukushima \\    
1958--1964  \, W. Fricke                   & 1985--1988  \, B. L. Morando       &  2009--2012 \, G. H. Kaplan \\      
1964--1967  \, G. M. Clemence       & 1988--1991  \, P. K. Seidelmann &  2012--2015  \, C. Y. Hohenkerk \\
\hline
\end{tabular}
\end{center}
\smallskip

The final meeting of Commission~4
was held at the XXIXth IAU General Assembly (GA) in Honolulu in 2015.   At that time, the members of the commission decided that it would be appropriate to present within this volume some reflections on its history,
which extends over almost a century of enormous scientific progress.   In the following, we trace some of the scientific
issues that the commission has addressed over that time (usually in partnership with other commissions)
and some historical notes about the publications of the national institutions that have been active in
Commission~4's work.

\section{Scientific Issues}

\subsection{Reference Systems \, {\it (G. H. Kaplan)}}

Any kind of ephemeris work relies on well-defined coordinate systems.
The very first resolution passed by Commission~4 at the first IAU GA (Rome, 1922) was an endorsement of a
resolution from Commission~8 (Meridian Astronomy) calling for the directors of the national ephemeris 
offices to adopt a ``uniform system of Standard Star places'' --- essentially, what we would now call
a standard astronomical reference frame.

The terminology that has become commonly used in the astrometric and
geodetic communities distinguishes between a {\it reference system\/} and a {\it
reference frame\/}.  A {\it reference system\/} is the complete
specification of how a celestial or terrestrial coordinate system is to be
formed.  \NOPRINT{Both the origin and the orientation of the 
fundamental planes (or axes) are defined.  A reference system
also incorporates a specification of the fundamental models
needed to construct the system; that is, the basis for the algorithms used to
transform between measurable quantities and reference data in the system.}
A {\it reference frame\/}, on the other hand, consists of a set of identifiable fiducial
points on the sky --- or on the Earth --- along with their coordinates and rates of change,
which serves as the practical realization of a reference system. 

For most of the 20th century, the fundamental plane of 
astronomical reference systems was the extension of the Earth's
mean equatorial plane, at some date, to infinity.  The mean equinox at that date defined the azimuthal origin
along that plane, i.e., the origin of right ascension.  The standard astronomical reference frames
were based on fundamental star catalogs compiled mainly from transit telescope
observations of bright stars, planets, and the Sun.
The observations of solar system objects established the location of the equator and
equinox with respect to the stars.   
The orientations of these reference frames were therefore closely tied to the quality of the
planetary ephemerides available at the time.  Conversely, improvements
to the planetary ephemerides depended --- until the space age --- on the accumulation of transit
telescope observations used to establish these reference frames.  Before 1997, the fundamental
reference frames that had been endorsed by the IAU were, first, the 1925 Eichelberger list of standard stars from the US Naval
Observatory in Washington, then the FK3, FK4, and FK5 star catalogs from the Astronomisches Rechen-Institut in Heidelberg. 
See the introduction to the FK4 \citep{fricke63} and \citet{eichhorn74} for reviews.

The development of very long baseline radio interferometry (VLBI) as an astrometric and geodetic 
technique in the 1970s and 1980s changed the thinking on how astronomical reference frames
should be constructed.  Not only were the VLBI-determined coordinates of radio sources
very precise, but the ensemble of extragalactic objects that were observed formed a more
rigid frame than one defined by stars, and which was not subject to spurious rotations. 
VLBI observations were quite sensitive to the motions of the equator (hence they also provided
excellent determinations of precession, nutation, polar motion, and UT1) but had no tie to the
equinox, because no natural solar system objects could be observed.  The development of
space-based optical astrometry, which has no fundamental tie to either the equator or equinox,
also motivated new thinking about astronomical reference frames generally.  The histories of VLBI  and of space-based optical astrometry have been reviewed respectively by \citet{clark03} and \citet{hog11}.

Beginning in 1988, a number of IAU working groups began
considering the requirements for modern astronomical
reference systems.  The resulting
series of IAU resolutions, passed at the XXIst, XXIInd, XXIIIrd, and XXIVth IAU GAs (Buenos Aires, 1991;
The Hague, 1994; Kyoto, 1997; and Manchester, 2000)
effectively form the specifications for 21st century astronomical reference
systems and the frames that realize them.  The directions of the axes of the {\it International
Celestial Reference System} (ICRS) are defined by the adopted positions of a specific set of
extragalactic objects, which are assumed to have no measurable
proper motions (although variations in source structure can mimic proper
motions).  The ICRS axes are consistent, to about 0.02
arcsecond, with the mean equator and equinox of J2000.0.  However, the ICRS axes are
meant to be regarded as fixed directions in space that have an existence
independent of the dynamics of the Earth or the particular set of
objects used to define them at any given time.

The ICRS is realized at radio wavelengths by 
the {\it International Celestial Reference Frame} (ICRF), which is now
in its second release (ICRF2), with a third (ICRF3) planned by 2018.  These
are all VLBI catalogs of compact extragalactic radio sources, many of which have faint optical
counterparts, typically $m_{\rm v} \gg 18$.

The ICRS is realized at optical wavelengths by a subset of stars in the
Hipparcos Catalogue, which was the result of a European Space Agency (ESA) space astrometry mission
launched in 1989.   The subset, referred to as the {\it Hipparcos Celestial Reference Frame} (HCRF),
includes stars as faint as $m_{\rm v} = 12$ with
uncomplicated and well-de\-ter\-mined proper motions (e.g., no
known binaries).   ESA's current space astrometry mission, Gaia, now conducting observations,
will provide a very high accuracy replacement for the HCRF in the early 2020s.  

Modern high-precision planetary ephemerides such as those from the
Jet Propulsion Laboratory (US), Institute of Applied Astronomy (Russia),
and the Institut de M\'{e}canique C\'{e}leste et de Calcul des \'{E}ph\'{e}m\'{e}rides (France)
are referred to the ICRS using VLBI observations of
interplanetary spacecraft made with respect to ICRF sources.  Although the equator and equinox
no longer define fundamental directions, knowing their changing location within the ICRF is still
necessary for many practical applications; precession and nutation models still describe their motions.

A relativistic framework (metrics, potentials, etc.) for geocentric and solar system barycentric reference systems was recommended
by resolutions passed at the XXIst and XXIVth IAU GAs (Buenos Aires, 1991, and Manchester, 2000).   

For a more complete narrative, see Chapter~4 of \citet{ES12} or Sections 9.2 and 9.4 in \citet{kopeikin11},
and the references therein.

\subsection{Time Scales \, {\it (P. K. Seidelmann)}}

Defining the most appropriate independent argument for ephemerides has been a continuing subject of discussion and change for most of the history of Commission~4.

Mean solar time, based on the rotation of the Earth, was the (supposedly) uniform time scale introduced by Ptolemy in about 150 AD; it provided a civil and scientific time scale for 18 centuries.  In the late 1800s there were suspicions from lunar observations that the rotation of the Earth varied. In the 1920s, de Sitter and Spencer Jones proved the variability from lunar and planetary observations --- systematic residuals in ecliptic longitude were proportional to the mean motions, indicating a problem with the common time scale.  For a uniform time scale, Ephemeris Time (ET) was proposed by Clemence and adopted by the Xth IAU GA (Moscow, 1958).  The ephemeris second was defined to be the mean solar time second of 1900.0, but ET was determined in practice from the motion of the Moon.

In the 1950s, atomic clocks became available, and an atomic time scale, TAI, began in 1954.  Markowitz determined the length of the ephemeris second in terms of the cesium atom frequency, thus defining the SI second, used as the unit for atomic time.  \NOPRINT{Ephemeris Time was difficult to determine from observations, so in 1968 International Atomic Time (TAI) replaced Ephemeris Time as a time scale after 1954.} 

The modern implementation of mean solar time, based on observations of the rotation of the Earth, is Universal Time, specifically UT1.  UT1 and its rate of change (length of day) remain important quantities in geophysics, astronomy, and navigation; for example, UT1 is required to obtain sidereal time.  However, in the modern context, UT1 is probably better considered a measurement of the Earth's rotation angle, in time units, rather than as a time scale.  \NOPRINT{UT2, no longer commonly used, is obtained from UT1 by removing periodic terms corresponding to the known annual and semiannual variations in length-of-day.}  The international civil time scale, Coordinated Universal Time, UTC, is based on atomic time (i.e., the SI second on the geoid) but is modified by leap-second adjustments to remain within 0.9\,s of UT1.  By convention, UTC is an integral number of seconds offset from TAI.     \NOPRINT{The raw observed time is labeled UT0. When corrected for polar motion, it is UT1, and when corrected for annual variations, it is UT2. For a period UT2 was used, but then UT1 came into common use for UT.  In 1972, Coordinated Universal Time (UTC) was introduced, based on TAI and with leap seconds to maintain UTC within 0.9 second of UT1.}  Values of $\Delta$UT1 = UT1--UTC are now regularly distributed by the International Earth Rotation and Reference Systems Service (IERS).

By the 1970s, fundamental planetary ephemerides were being computed that used some form of {\it coordinate time}, in the context of general relativity, as the independent argument.  This approach influenced subsequent IAU recommendations.
Resolutions passed at the XVIth and \hbox{XVIIth} IAU GAs (Grenoble, 1976, and Montreal, 1979) recommended two new time scales: one for geocentric ephemerides, named Terrestrial Dynamical Time (TDT), and one for barycentric ephemerides, named Barycentric Dynamical Time (TDB).  TDT was redefined and renamed in 1991 to Terrestrial Time (TT).  The long-term rates of both TT and TDB are based on the SI second on the geoid (they differ in small periodic terms), and TDT/TT is considered to be continuous with ET.  In 1991, the IAU introduced Geocentric Coordinate Time (TCG) as the coordinate time scale for a geocentric reference system, and Barycentric Coordinate Time (TCB) as the coordinate time scale for a solar system barycentric reference system.  These coordinate systems, their time scales, and the relationships between them were specified more precisely by IAU resolutions passed at the XXIVth IAU GA (Manchester, 2000).  TT and TDB remain in common use, and IAU resolutions passed in 2000 and 2006 now define these two time scales by linear expressions with fixed coefficients (defining constants) that relate them to TCG and TCB, respectively.    

The difference $\Delta$T = TT--UT1 (formerly ET--UT) is studied to determine the history of the rotation of the Earth.

See \citet{mccarthy09}, Chapter~3 of \cite{ES12}, and Section 9.3 in \citet{kopeikin11} .

\subsection{Astronomical Constants \, {\it (P. K. Seidelmann, G. H. Kaplan)}}

The Paris Conference of 1896, Newcomb's subsequent work on precession, and Brown's lunar theory established many of the fundamental astronomical constants that were widely used for most of the 20th century.  These constants are the basis for many of the models and calculations used by Commission~4 members and institutions.  Except for a Commission~4 resolution at the VIth IAU GA (Stockholm, 1938) regarding the Gaussian gravitational constant, the IAU did not become involved in establishing a system of fundamental astronomical constants until after World War~II.  A resolution at the VIIIth IAU GA (Rome, 1952) reaffirmed the conventional values of the standard list of constants (without listing them).  The situation up to 1960 has been well described in Chapter~6 of the 1961 {\it Explanatory Supplement to the A.E.} \citep{ES61}. 

Following an IAU symposium in Paris in 1963, a working group was established, chaired by Walter Fricke, to review the system of constants and recommend changes.  At the XIIth IAU GA (Hamburg, 1964), an IAU System of Astronomical Constants was approved for introduction into the national ephemerides beginning in 1968. The main changes were in the values of the astronomical unit, aberration, figure of the Earth, geocentric gravitational constant, and the Earth/Moon mass ratio. Except for the lunar ephemeris and the day numbers, the ephemerides were not changed, because the changes were so small. The ephemeris of the Moon was corrected, based on the corrections to the constants and term 182 in the lunar theory.  The precessional constants and the motion of the ecliptic were not changed --- Newcomb's values were still used.  See the ``Supplement to the A.E. 1968'' in the 1974 reprint of the 1961 {\it Explanatory Supplement to the A.E.}

Immediately prior to the XIVth IAU GA (Brighton, 1970), an IAU colloquium was held in Heidelberg, Germany, to consider the future of the IAU System of Astronomical Constants and the inclusion of the theory of relativity in ephemeris work.  A plan for the replacement of Newcomb's system was developed, with a proposed schedule for adoption in 1976 and introduction into use in 1984.  Working groups were developed for precession, ephemerides, and constants. The working groups corresponded, met, modified their memberships, and reached agreements. At the XVIth IAU GA (Grenoble, 1976), most of the constants, ephemerides, reference frame, and time scales were agreed upon. A new standard epoch, J2000.0, was approved.  The list of constant values is referred to as the IAU (1976) System of Astronomical Constants.  The values of the equinox offset and motion were left for a final determination by Walter Fricke as part of the analysis for the forthcoming FK5 fundamental star catalog. 

Subsequently, a Working Group on Nutation was established (see Section 1.8).  At the XVIIth IAU GA (Montreal, 1979) the names of the new time scales and a new nutation theory were adopted.  The Jet Propulsion Laboratory in the US produced a planetary and lunar integration, DE200/LE200, to serve as the basis for new planetary and lunar ephemerides consistent with the new reference system.  In 1981, {\it The Astronomical Almanac} was first published, replacing the previously separate British and US astronomical ephemeris publications; it was in a revised format, in preparation for the new reference system in 1984.  Also in 1981, Fricke's values of the equinox offset and motion to be used in the FK5 were selected.  The 1984 and later editions of the national ephemeris publications were based on the new system, although the FK5 star catalog was not published until 1988.  Simon Newcomb's system of constants and theories, which had been in use for 84 years, was finally replaced.  See \citet{kaplan81}.

The IAU (1976) System of Astronomical Constants remained in place for 33 years.  In 2006, the IAU established, within Division I (now A), a Working Group on Numerical Standards, which recommended a new set of constants.  These were adopted at the XXVIIth IAU GA (Rio de Janeiro, 2009) as the IAU (2009) System of Astronomical Constants \citep{luzum11}.  The Working Group has continued, and it regularly updates a separate online list of ``best estimates'' for constants whose values improve due to better observational data.  The XXVIIIth IAU GA (Beijing, 2012) changed the length of the astronomical unit in meters from an estimated value to a defining constant (similar to the speed of light) whose value will not change (see Section~1.7).

\subsection{General Versus Special Perturbations --- Analytic Theories and Numerical Integrations \, {\it (P. K. Seidelmann, A. Fienga, E. V. Pitjeva)}} 

Prior to the advent of mainframe computers, the computation of ephemerides involved analytic theories, also called ``general theories'' or ``general perturbations,'' based on a series expansion of the disturbing (perturbation) function.  Such a development results in a long Fourier series of trigonometric terms that could be evaluated for specific times to form an ephemeris.  Until 1960, the theories of LeVerrier, Newcomb, and Hill were generally adopted for the planets, along with Brown's theory of the Moon.  Determining the theories and improving them was an extensive effort.  Punched card equipment and computers were introduced for evaluating the theories to determine ephemerides for the almanacs.  

With the availability of mainframe computers after World War II, it became possible to perform numerical integrations, also called ``special perturbations,'' that could be fit to observations and recomputed for improved accuracy.  The first large-scale N-body integration by computer, {\it Coordinates of the Five Outer Planets}\/ by Eckert, Brouwer, and Clemence, was published in 1951; each integration step took ``less than two minutes'' on the vacuum-tube computer used \citep{eckert51}.  In the US in the 1960s, the Jet Propulsion Laboratory (JPL) and the Massachusetts Institute of Technology (MIT) developed planetary integration programs to provide the improved accuracies required for the reduction of radar observations and planned planetary space missions.  The approach is described by \citet{newhall83}.  The accuracy of the new observation types, improvements in the knowledge of the dynamics of the solar system, and the increased precision of the computations provided a means to test the effects of general relativity in the solar system.  However, the Apollo manned lunar missions in the 1960s used the Improved Lunar Ephemeris, which was based on Brown's theory of the Moon.

In the 1960s in the US,\NOPRINT{Lloyd Carpenter introduced the use of Chebyshev polynomials for general theories and for interpolating ephemerides;  Chebyshev polynomials are still used in all the modern ephemeris developments.} L.~Carpenter and P.~K.~Seidelmann developed software to generate planetary general theories by computer.  P.~Bretagnon, J.~Chapront, and M.~Chapront-Touze initiated the general theory activity in France.  While general theories could be generated by computer, they required very long series, and better accuracies could be achieved with numerical integrations. In 1984, a numerical integration of the planets and Moon (DE200/LE200) by E.~M.~Standish and collaborators at JPL were introduced in British and American almanacs, as well as in the Russian Astronomical Yearbook.  More advanced integrations by JPL were introduced in the British and American almanacs in 2003 and 2015. 

In France, the tradition of the analytical theories continued into the 2000s with the analytical solutions of the secular motion of the planets. The first comparisons of spatial observations to these solutions, describing the motion of planets by means of Poisson series, quickly showed the limit of the method; the accuracy of the modern observations would have required a significant increase of the length of the series. In 2003, it was decided to develop new numerical ephemerides together with maintaining the analytical solutions. The results of two recent INPOP (Int\'{e}grateur Num\'{e}rique Plan\'{e}taire de l'Observatoire de Paris)
integrations by the Institut de M\'{e}canique C\'{e}leste et de Calcul des \'{E}ph\'{e}m\'{e}rides (IMCCE) are represented in Table~1.

In Russia, since the 1970s, precise planetary ephemerides to support Russian
space flights have been constructed at the Institute of Applied Mathematics,
at the Institute of Radioengineering and Electronics, and at the
Institute of Theoretical Astronomy (ITA). At ITA, the first planetary
theories (AT-1, 1978) were analytical, created by the group headed by
G.~A.~Krasinsky.  Krasinsky developed an effective algorithm for the expansion of the
perturbation function based on its representation as the sum of products of some special functions. However, results of the comparison between the AT-1 theory and the first computed numerical
integration showed that the latter matched observations noticeably
better. Since the 1980s, only numerical planetary ephemerides have been
developed at ITA and then at the Institute of Applied Astronomy (IAA),
where the series of EPM  (Ephemerides of Planets and the Moon) integrations has been
produced. These ephemerides were introduced in Russian Astronomical Yearbook in 2006.

General theories also provided a basis for long-term studies of the stability and chaos in the solar system. Wisdom and Laskar have led efforts concerning such studies.

\begin{table}
\begin{center}
\small
\begin{tabular}{l l | r r | r r | r r | r r}
\hline
\multicolumn{2}{l}{ } & \multicolumn{2}{c}{Le Verrier } & \multicolumn{2}{c}{DE200} & \multicolumn{2}{c}{DE421, INPOP08,} & \multicolumn{2}{c}{DE430, INPOP15a,}\\
\multicolumn{2}{l}{ } & \multicolumn{2}{c}{} & \multicolumn{2}{c}{} & \multicolumn{2}{c}{EPM2008} & \multicolumn{2}{c}{EPM2014}\\
\multicolumn{2}{l}{ } & \multicolumn{2}{c}{\it c. 1900} & \multicolumn{2}{c}{\it 1980} & \multicolumn{2}{c}{\it 2008} & \multicolumn{2}{c}{\it 2014--2015}\\
 & & angle & distance & angle & distance  & angle & distance & angle & distance  \\
 & & $''$\D   &  km   & $''$\DD    &  km     & $''$\DDD   &  km  & $''$\DDD  &  km   \\
 Mercury & &1\D\w{.}   &   450 &  0.020 &     5 &   0.0030 &  0.40 &  0.0002 & 0.020 \\
Venus   & & 0.5 &   100 &  0.020 &     2 &   0.0004 &  0.02 &  0.0002 & 0.004 \\
Earth   & &     &       &  0.010 &     1 &   0.0004 &  0.01 &  0.0002 & 0.002 \\
Mars    & & 0.5 &   150 &  0.010 &     3 &   0.0004 &  0.01 &  0.0002 & 0.002 \\
Jupiter & & 0.5 &  1400 &  0.1\DD   &    50 &   0.0050 &  2.\DD   &  0.0040 & 1.5\DD   \\
Saturn  & & 0.5 &  3000 &  0.2\DD   &   350 &   0.0010 &  0.2\D  &  0.0002 & 0.2\DD   \\
Uranus  & & 1\D\w{.}   & 12700 &  0.4\DD   &  5000 &   0.0100 &  100.\DD &  0.0050 &  50.\DDD  \\
Neptune & & 1\D\w{.}   & 22000 &  1.0\DD   &  8000 &   0.0100 &  300.\DD &  0.0050 & 200.\DDD  \\
Pluto       & &      &               &  2.5\DD   & 80000 &   0.0200 & 1200.\DD &  0.0200 & 500.\DDD  \\
\hline
\end{tabular}
\caption{\small The increasing accuracy of planetary positions in ephemerides from the beginning of the 20th
century.   For the last three columns, the uncertainties are for J2000.0; for the inner planets through Saturn, the angular error is the maximum uncertainty in right ascension or declination over the synodic period centered on J2000.0.  The modern US, French, and Russian ephemerides are similar in overall accuracy.  Note that the uncertainties can be computed in different ways, which can lead to somewhat different accuracy estimates.}
\normalsize
\end{center}
\end{table}

Table 1 shows the evolution of the accuracy of planetary ephemerides. The one published by Gaillot (named after Le Verrier) at the start of the 20th century is an example of what could be expected for an analytical solution of planetary motions before the age of modern observations and numerical integrations.

\subsection{Evolution of the Planetary Ephemerides in the Space Age \, {\it (W. Folkner)}}

Prior to the space age, the planetary ephemerides were estimated primarily from astrometric observations of planets and planetary satellites with accuracy limited to about $1''$ from troposphere fluctuations, in addition to systematic effects in star catalogs used to define the celestial reference frame. With the launch of spacecraft to the Moon and planets came the ability to greatly improve the ephemeris accuracy. Improved ephemeris accuracy in turn enabled more accurate spacecraft navigation for improved targeting and science observations.

The first spacecraft to the Moon and planets made flybys. Analysis of the Doppler shift of radio signals during the flybys allowed accurate estimation of the spacecraft positions with respect to the Moon or planet.  Radio range measurements from the spacecraft could then be used to determine the distance from Earth. This provided much higher accuracy than earlier planetary radar data, since the spacecraft positions were determined relative to the center of mass with greater accuracy than knowledge of planetary topography. However, the flyby data provided only limited geometric coverage.

The deployment of optical retro-reflector arrays on the Moon starting in 1969 provided a major step forward in measurement accuracy (see Section~1.6). Lunar laser ranging (LLR) measurements enable accurate estimation of the orbits of Moon, Earth, and Sun relative to the dynamical equinox and of Earth relative to Sun.  LLR is sensitive to lunar librations and Earth precession and nutation. Thus ephemerides determined by LLR are estimated with respect to the dynamical Earth pole and equinox, but not with respect to star catalogs.

Following initial surveys of the planets from spacecraft flybys, more in-depth surveys were performed by spacecraft in orbit about, or on the surface of planets, providing robust time series of measurements useful for ephemeris determination. This started with the Mariner~9 spacecraft in orbit about Mars in 1971 followed by the Viking orbiters and landers in 1976. The Viking~1 lander in particular provided seven years of radio range measurements to a fixed point on the surface of Mars with accuracy of ~5 m. These data improved the accuracy of the orbit of Mars relative to Earth better than $0.01''$.

Along with LLR, very-long baseline interferometer (VLBI) measurement accuracies improved rapidly in the 1970s and 1980s with observations of extra-galactic radio sources used to determine a celestial reference frame and station location and Earth orientation with respect to that frame. Initially the VLBI frame was defined as best possible with respect to the best star catalogs that in turn were aligned to the Earth's equator and equinox. However, uncertainties in the star catalogs and ambiguities in the definition of the equinox limited the accuracy of the VLBI frame.  By the late 1980s, Earth rotation parameters were being determined with respect to a nominal set of radio source positions. 

Since the 1990s, VLBI measurements with respect to what we now call the ICRF (see Section~1.1) have formed the basis for regular measurement of Earth orientation, augmented by other techniques including satellite laser ranging, GPS, and DORIS (satellite doppler). The intensive program for measurement of Earth orientation and station locations allowed determination of relative station locations from different techniques with accuracies of a few millimeters. Comparison of station locations and estimated Earth orientation between VLBI and LLR methods in 1994 allowed determination of the orientation of the ICRF with respect to the dynamical Earth equator and equinox with accuracy of $0.005''$. The ICRF was determined to be offset from the dynamical equator and equinox by $0.04''$. The ICRF was not changed as a result, but the offset is now taken into account by applying a ``frame bias'' matrix in conjunction with the Earth precession and nutation models.

The improved accuracy of Earth and Mars orbits with respect to the ICRF and the consistency of Earth orientation measurements enabled improved targeting of a series of space missions that has led to a continuous set of measurements of the Martian orbit. The first of these was Mars Pathfinder in 1997 that landed on Mars directly on arrival from Earth, rather than first going into orbit about Mars. This was followed by a series of Mars orbiters in low-altitude orbits starting with the Mars Global Surveyor in 1999 and continuing to the present, with five active orbiters: Mars Odyssey, Mars Reconnaissance Orbiter, MAVEN, Mars Express, and the Mars Orbiter Mission. The positions of these orbiters are now determined with respect to the center of mass of Mars with accuracy better than 1~m. This allows radio range measurements to the orbiters to provide a continuous series of measurements of the Earth-Mars distance with meter accuracy, giving very accurate determinations of the orbits of Earth and Mars. Mars is near the main asteroid belt so its orbit is significantly perturbed, especially by the largest asteroids and asteroids near resonance with the martian orbit. The continuous series of range measurements to Mars orbiters has allowed improving estimates of the mass parameters ($GM$) of the major perturbing asteroids that results in continuing improvement in prediction of the Mars orbit.

In order to enable more accurate landing on Mars, a series of VLBI measurements of orbiting spacecraft with respect to radio 
sources known within the ICRF has been performed to better estimate the orientation of the orbits of Earth and Mars with respect to the ICRF. With improvements in measurement techniques over the past two decades, the current measurement accuracy has allowed the determination of the orbit orientation to $0.00025''$, corresponding to knowledge of the position of Mars with 200 m accuracy.  The figure below shows the improved accuracy in the orbit of Mars with respect to the ICRF over the past 25 years resulting from the series of spacecraft range and VLBI measurements.\\

\begin{center}
\includegraphics[width=5.0in,trim=0.0in 0.0in 0.0in 0.0in,clip=true]{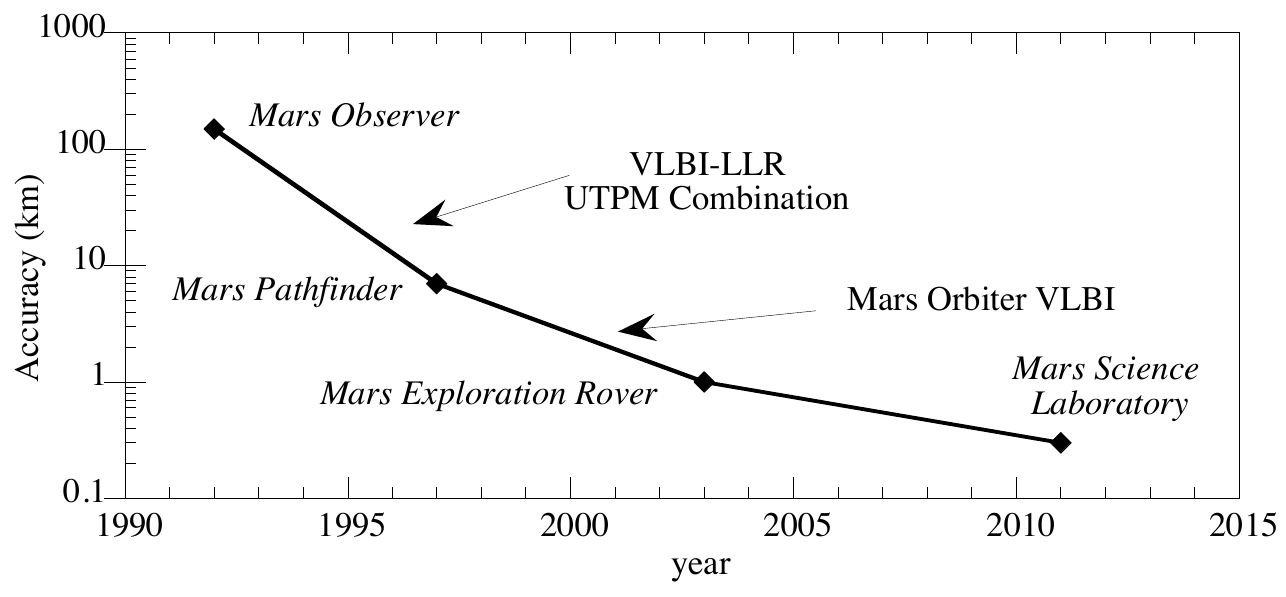}
\end{center}

The orbits of Venus and Mercury achieved similar accuracy with the Venus Express and MESSENGER orbiting missions. Radio range measurements to these orbiters over several years allows accurate estimates of the Venus and Mercury orbits with respect to Earth, and tied to the ICRF through the VLBI measurements of Mars orbiters. (The earlier Magellan orbit about Venus did not have ranging capability so did not contribute much for the Venus ephemeris improvement.)

The orbit of Saturn is now known with accuracy comparable to that of the inner planets through radio tracking of the Cassini spacecraft since it entered orbit about Saturn in 2004. The accuracy of the orbit of Jupiter is not known with similar accuracy due to lack of any orbiters allowing a time series of range and VLBI measurements. The orbit of Jupiter is now estimated mainly from radio tracking of seven spacecraft that flew past Jupiter on their way to the outer solar system. While the Galileo spacecraft did orbit Jupiter for several years, the failure of its high-gain antenna precluded radio range and high-accuracy VLBI measurements. The Jupiter orbit is expected to be improved when the Juno orbiter arrives at Jupiter in 2016.

The orbits of Uranus, Neptune, and Pluto (now a dwarf planet) are still primarily determined from astrometric measurements, with much less accuracy than for the other planets. These may be improved over the next several years by astrometric measurements from the Gaia mission.

\subsection{The Lunar Ephemeris \, {\it (J. G. Williams)}}

Prior to the first spacecraft, the lunar ephemeris was realized with an analytical theory with fits of the orbit parameters based on optical astrometry: meridian transits and stellar occultations. Early spacecraft flying by, around, and to the Moon included Zonds, Rangers, Lunar Orbiters, Surveyor and Luna landers, and Apollo manned landings. As a counterpart to numerical integrations of spacecraft trajectories for navigation at the Jet Propulsion Laboratory, numerical integration of the orbits of the Moon and planets was initiated. Early radar ranges to the surfaces of Venus, Mars, and Mercury marked the beginning of observations with better accuracy than optical astrometry. 

During these early years, physical librations of the synchronous Moon were represented by analytical series. The differences in the lunar moments of inertia were determined from heliometer measurements of orientation. One 3-year longitude libration term is near a resonance. 

During 1969--1973, five laser retroreflector arrays were placed on the Moon: three by the Apollo astronauts and two French-built arrays on the Soviet Lunokhod rovers carried by Luna landers. Lunar laser ranging (LLR) was achieved at Lick observatory two weeks after the Apollo~11 landing in 1969; a long duration ranging program was started at McDonald Observatory a few weeks later. Early meter-accuracy ranges were quickly improved to a few decimeters by 1970 and to 0.2~m by 1972. Occasional ranges were acquired in the Soviet Union. Lunar ranges were acquired at Orroral, Australia. The rapid three orders-of-magnitude improvement in observation accuracy from hundreds of meters, before ranging, to decimeters was a challenge for data analysis. It wasn't only the lunar ephemeris that needed improvement, but also observatory coordinates, Earth orientation, retroreflector locations, and lunar physical librations. 

The 1970s saw a vigorous effort to improve the modeling of the lunar ranges. By 1973 the root-mean-square (rms) residual was 3~m, still an order of magnitude larger than the range accuracy. Two groups discovered that physical librations were sensitive to the third-degree gravity field. In 1976, the first LLR test of the equivalence principle was published; also free physical librations were detected. The rms residual was 0.4~m in 1976, about twice the data accuracy. The year 1978 saw the first definitive LLR publications of the tidal acceleration of the Moon.  In the 1970s, several papers had found larger (negative) accelerations than the 1939 value of $-22.4''$/cy$^2$ from Spencer Jones, but the 1978 LLR value came out to be $-24''$/cy$^2$. Also concordant, a 1975 value of $-26''$/cy$^2$ used Mercury transits of the Sun to establish the time scale. The LLR observations were referred to atomic time while the 1939 value used Earth rotation as a clock. The LLR results did not support any distinction between atomic time and time in the gravitational equations of motion. 

In the early 1980s, physical librations were added to the simultaneously integrated orbits of the Moon and planets. A new McDonald laser ranging system (MLRS) replaced the 2.7~m operation, ranging started at the Observatoire de la C\^{o}te d'Azur, France, and at the Haleakala Observatory, Hawaii. By the end of the 1980s, the range accuracy was less than 5~cm and LLR had been used to test the equivalence principle and relativity. 

In 1991, a correction from LLR to the IAU-adopted precession of the equinox was published, followed soon after by a compatible result from VLBI. By 1995, the range accuracy was 2~cm. Analysis improved the value of the obliquity and detected the Love number of the Moon and its tidal Q. Later, the obliquity rate was determined. In 2006, the Apache Point Observatory, USA, began a ranging program. Designed for millimeter ranging, this data is a new challenge to analysis. 

There are now (in 2015) 20,200 lunar laser ranges that have been accumulated. Two stations have vigorous ranging programs: Observatoire de la C\^{o}te d'Azur, France, and Apache Point Observatory, USA. Ranges for the last few years are fit with an rms residual of 1.5~cm.  Since the beginning of the LLR decades, the lunar distance has been known to a fraction of a meter. The tidal acceleration is established as $-25.8''$/cy$^2$ and the semimajor axis rate is +38.1~mm/yr. Physical librations, including free librations, are determined to a few milliarcseconds. Analysis centers exist in France, Germany, Russia, and the United States. The LLR experiment has made contributions to ephemerides, astronomical constants, terrestrial geodesy, lunar science, and gravitational physics.

\subsection{The Astronomical Unit \, {\it (E. V. Pitjeva, A. Fienga)}}

The astronomical unit (au) --- approximately, the average distance of the Earth from the Sun --- is
one of the basic units of astronomy,
determining the scale of the solar system and anchoring the entire system of astronomical distances.
Classical measurements of the positions of objects in the solar system were angular.
A period ($P$) or the mean motion $n=2\pi/P$ of a solar system body
was much more easily measured than its distance; distances were specified in au, whose length in
ordinary linear units was not well known.  In astronomy, the adopted units were those of the solar mass ($M$),
the mean solar day ($d$), and the Gaussian (solar) gravitational constant ($k$ = 0.01720209895).
The au is then the unit of length which is consistent with these units 
by Kepler's third law: $n^2a^3 = k^2M$, where $a$ is the semimajor axis of the orbit
(around the Sun) and $M$ is the mass of the Sun.  For a massless particle  
at 1~au from the Sun in Keplerian motion, we have $a$=1 au, $M$=1~$M_{\odot}$, 
so the mean motion is $n=k$ and the period
$P= 2\pi/k =365.2568983...$ days; the latter is known as a Gaussian year,
which is based on old measurements of the Earth's mean motion.

The term ``astronomical unit'' appears at the beginning of the twentieth century.
The VIth IAU GA (Stockholm, 1938) adopted a resolution fixing $k$ at the above value; this was reiterated in 
the lists of astronomical constants adopted by the IAU in 1964 and 1976 (see Section 1.3).
In the IAU (1976) System of Astronomical Constants we find the au defined as
``The astronomical unit of length is that length ($A$) for which
 the Gaussian gravitational constant ($k$) takes the value of 0.017\,202\,098\,95
 when the units of measurements are the astronomical unit of length, mass and
 time. The dimensions of $k^2$ are those of the constant of gravitational ($G$),
 i.e., L$^3$M$^{-1}$T$^{-2}$.''  (Resolution of the XVIth IAU GA (Grenoble, 1976))

The first attempt to estimate the length of what we now call the au was made by
Aristarchus of Samos in the third century BC.  In the late 1800s,
an accuracy of 0.1$\%$ was obtained from the geocentric parallaxes of asteroids.  In 1950,
the astronomer Rabe achieved an accuracy of 0.05$\%$ ($\pm$65\,000~km) from 
the close approach of Eros to the Earth \citep{rabe50}.
All such determinations were made by optical triangulation methods
with large scatter.  Moreover, the ranges of the probable errors did not overlap.

In 1961, the first successful radar ranging to Venus was obtained, and the first 
estimates of au values from radar had accuracies from 2000 km to 130 km.
However, a discrepancy between the new radar determination 
of the au and the photographic value by Rabe remained.
\citet{lieske68} collected 8639 observations of Eros, from 1893 to 1966, reduced them, 
adjusted the ephemerides, and obtained an au value closer to that from radar.
Subsequently, the value of the au was determined exclusively from radar ranging of planets
and spacecraft tracking.  The typical uncertainties in the value of the au decreased to
the level of a few meters.

The astronomical unit
is connected to the heliocentric gravitational constant by the definition given above. 
If one uses the classical astronomical units of au, day, and solar mass, the value of the
heliocentric gravitational constant, $k$, is fixed and the distance scale (semimajor axes of the
planetary orbits in au) is adjusted according to the observations.  Range observations allow for
determining the length of the au in meters.
In SI units, either of these values (au or $GM_{\odot}$) could be estimated by fitting ephemerides
to observations directly (but not both simultaneously), or the value of one of them could be obtained from
the value of the other.  A 2009 value of
au=149\,597\,870\,700~($\pm$3)~m by Pitjeva and Standish was consistent with a contemporary value
$GM_{\odot}$=1.327\,124\,400\,41\E{20}~($\pm$1\E{-10}) m$^3$s$^{-2}$ by Folkner.  These estimates
were obtained from the observational fit to ephemerides computed at IAA in Russia and JPL in the US. 
These values, along with other modern values of astronomical constants,
were accepted by the XXVIIth IAU GA (Rio de Janeiro, 2009) as part of the IAU (2009) System of Astronomical Constants (Resolution~B2; see Section 1.3 and \citet{luzum11}).

Beginning around 2010 at JPL in the US, IMCCE in France, and IAA in Russia, experiments began in fixing
the value of the au when the ephemerides were computed.  The value of $GM_{\odot}$ of the Sun was then
fitted to the observations along with the initial conditions for the planetary orbits and other adjustable ephemeris parameters.

The astronomical system of units (au, days, and solar masses) is convenient for many aspects of dynamical astronomy.
It is the astronomical system that is now used by the majority of users of planetary ephemerides (other than for navigation) 
and all the almanacs.   However, the classical definition of the astronomical unit is difficult
to comprehend and frequently leads to misunderstanding and misstatements; for example, the au is not the
semimajor axis of the Earth's orbit (which is 1\E{-6} greater).  Furthermore, it is desirable to have a
connection to the SI system of units used in most other fields of science.   For this purpose, we can define the astronomical day as simply 86\,400 SI seconds and the astronomical unit as some fixed number of meters.   The solar mass
is more problematic because in astronomy we never measure masses in kilograms (except for meteorites),
only mass ratios or the product $GM$, the latter now called the ``mass parameter''.  In addition, from solar physics and stellar evolution models we know that the solar mass is not constant but is likely decreasing.  Both $GM_{\odot}$ and its rate of change can now be solved-for parameters in the development of modern high-precision ephemerides \citep{pitjev13,fienga15}.

To make a better connection to SI units, Resolution B2 of the XXVIIIth IAU GA (Beijing, 2012) fixed the
astronomical unit of length at the value 149 597 870 700~m exactly, in agreement with the observationally obtained 
value adopted for the IAU (2009) System of Astronomical Units (see Section 1.3).  This definition (like that of
the astronomical day) is used with all time scales, such as TCB, TDB, TCG, TT, etc.  The length of the au in meters
thus has become a ``defining constant'' not subject to further changes, and the classical definition no longer applies.
The Gaussian gravitational constant $k$ has been removed from the list of astronomical constants.
The 2012 resolution also adopted ``au'' (lower case) as the conventional abbreviation for the astronomical unit.

Moreover, the value of the solar mass parameter, $GM_{\odot}$, unconnected with the au now, is
determined observationally in SI units.  At the XXIXth IAU GA (Honolulu, 2015), a resolution was passed that established a ``nominal'' value of $GM^N_{\odot}$  of 1.3271244\E{20}~m$^3$s$^{-2}$  for use in binary star and exoplanet work.  This value, which was deliberately set at a limited number of significant digits that will not change, could be
considered the third (mass) connection between the astronomical system of units and SI units, albeit a
low-precision one that depends on the value used for the fundamental gravitational constant, $G$.

\subsection{Non-Rigid-Earth Nutation Theory \, {\it (P. K. Seidelmann)}}

After the adoption of the IAU (1976) System of Astronomical Constants (see Section 1.3), the president of IAU Commission~4, V.~K.~Abalakin, in 1977 established a Working Group on Nutation. The members were V.~K.~Abalakin, H.~Kinoshita,  J.~Kovalevsky,  
C.~A.~Murray,  M.~L.~Smith,  R.~O.~Vicente,  J.~G.~Williams, Ya.~S.~Yatskiv, and P.~K.~Seidelmann (chair). The theory of nutation then in use was by E.~Woolard based on a rigid-Earth model \citep{wool53}.   The theory had several problems:  the constant of nutation
(the amplitude of the principal term in obliquity) 
was an empirical value and not consistent with the other astronomical constants; the effects of the Earth's non-rigidity were observationally significant;  determinations of UT1 and polar motion by modern observational methods were sufficiently accurate to be degraded by its use;  and the instantaneous axis of rotation of the theory rotated, relative to an Earth-fixed coordinate system, with a quasi-diurnal period.  Observational data indicated that the theory of nutation needed to be revised or replaced.

At the XVIIth IAU GA (Montreal, 1979), the Working Group recommended a rigid-body nutation series by Kinoshita with deformability given by Molodensky in 1961.  An unpublished nutation series by John Wahr (his 1979 Ph.D. dissertation) had just become available but had not been critically examined.  For astronomical applications, the differences between the two series were not then detectable.  The Working Group emphasized that the adoption was of nutation coefficients and not an endorsement of a particular Earth model.  However, in December 1979, the International Union of Geodesy and Geophysics (IUGG) adopted a resolution requesting the IAU to reconsider the choice of a nutation series. The IUGG objection was based on their interpretation that the IAU had implicitly endorsed the Molodensky Earth model.  Wahr's nutation was based on the Earth model 1066A of Gilbert and Dziewonski, which the IUGG considered to be the best Earth model then available. The IAU accepted the IUGG suggestion (by mail vote of the Working Group) and the Wahr model became the 1980 IAU Theory of Nutation.

The adopted nutation model was based on a non-rigid Earth model without axial symmetry, subject to tidal distortions, with a solid inner core, fluid outer core, but no oceans \citep{wahr81}. The constants were consistent with the IAU (1976) System of Astronomical Constants and in agreement with available observational data. The reference pole was selected so that there were no diurnal or quasi-diurnal motions of this pole with respect to either a space-fixed or Earth-fixed coordinate system.  Dynamical variation of latitude, also called forced diurnal polar motion, is included implicitly in the theory. The 1980 theory includes externally-forced motions of the Earth's rotation axis, but not free motions or complex phenomena, such as ocean tides, atmospheric winds, and currents in the ocean or core.  This nutation theory was based on previous work by Kinoshita and Gilbert and Dziewonski;  see \citet{seidel82}.

The Wahr nutation series was used for 20~years; for the most part, this ended Commission~4's leadership in this area.  After VLBI observations had covered an entire cycle of the principal (18.6-year) term of nutation, the XXIVth IAU GA 
(Manchester, 2000) recommended a change to the current nutation model of \citet{mathews02}.   A new working group is now considering even more sophisticated models.

\subsection{Pluto \, {\it (P. K. Seidelmann, J. Weratschnig)}}

After the successful discovery of Neptune in 1846, predicted by celestial mechanics and based on the residuals in the Uranus observational data, a number of studies were undertaken, particularly by William Pickering and Percival Lowell, into the positional residuals of Uranus and Neptune.  It is difficult to understand the methods in some of the published papers, but they predicted a number of additional planets.  Searches for Lowell's ``Planet X'' were conducted in the early 1900s at Lowell Observatory in Flagstaff, Arizona.  In 1930, about a year into a systematic photographic search, astronomer Clyde Tombaugh took plates near Lowell's predicted position for a planet and, when blinking them, found a slowly moving object.  Follow-up observations confirmed the discovery of a planet \citep{tomba97}. The discovery was announced on Lowell's birthday in 1930 and the object was named Pluto. The object had a period of 250 years and was 15th magnitude in brightness.  Pluto was initially largely ignored by Commission~4, but not by Commission~16 (Physical Observations of the Planets, Comets and Satellites).

Mass estimates for Pluto gradually decreased from the original value, close to that of the Earth, to about that of
Mars ($\approx$1/10 Earth mass) in the IAU (1976) System of Astronomical Constants.  There had always been
doubts that Pluto was the planet predicted by Lowell.   In 1978, Pluto's moon Charon was discovered on plates taken at the US Naval Observatory in Flagstaff \citep{christy97}.  This provided for the first time a reliable determination of the planet's mass; it was only about 1/400 Earth mass.  It became clear that the discovery of Pluto near its predicted position was serendipity.  Subsequent predictions of another
outer ``Planet X'' based on the residuals in the observational data of Uranus and Neptune were not confirmed, and the residuals mostly disappeared when improved planetary masses and ephemerides were determined \citep{standish93}.  Pluto kept its status as the ninth planet over the next decades --- although many astronomers thought that it was always the odd  one: There were rocky inner planets, there were gas giants, and then there was Pluto, far away in the outskirts of the solar system in a significantly eccentric and inclined orbit.  See also \citet{hoyt80}.

Improvements in detector technology changed our understanding of the outer solar system.  The application of charge-coupled device (CCD) detectors to astronomy in the 1970s and 1980s greatly improved imaging capabilities over those of photographic plates.  CCDs had a much higher quantum efficiency, were linear in response over a large dynamic range, and directly produced digital output.  The manufacturing process and quality of the CCDs, as well as the number of pixels per chip, improved significantly with time. By 1990, CCDs had effectively replaced photographic plates for optical astronomical observations and the latter were becoming difficult to obtain.  A specific result was the discovery of faint trans-Neptunian objects in the 1990s.  (CCDs also enabled the discovery of many small satellites of the outer planets.)  Several distant objects with estimated sizes comparable to that of Pluto were discovered.  One object discovered in 2005, later named Eris, was originally estimated to be more massive and also slightly larger than Pluto (Eris is now estimated to be slightly smaller than Pluto although its mass is greater).  This raised the obvious question: Is Eris a tenth planet, and if so, what about all the other, similar objects in that far region of the solar system?  If we decide they are not planets, what makes Pluto unique?  The status of Pluto as a major planet again came into question.
  
At the XXVIth IAU GA (Prague, 2006), the long discussed topic was brought to a decision.  The
main issue was the lack of an
agreed-upon definition of the term {\it planet}.  The IAU Executive Committee had formed a sub-committee to prepare a definition, which was submitted to the GA at its opening session.  It would have had the effect of increasing the number of planets in the solar system to perhaps~12.
This proposed definition was quickly abandoned, and the final IAU process was a remarkably democratic one, at least
for the participants of the GA. Many voices were heard, and the definition evolved over the next week.  In the end, two resolutions concerning this topic were passed.  Resolution 5a defined three main classes for bodies in our solar system (excluding satellites).  Pluto was thereby ``demoted'' to {\it dwarf planet}\/ status.  Resolution 6a introduced Pluto as a prototype for a new category of trans-Neptunian objects.

\section{Historical Notes on a Few Publications}

\subsection{History of Hv\v{e}zd\'a\v{r}sk\'a ro\v{c}enka --- Astronomer's Yearbook \, {\it (J. Vondr\'ak)}}

Hv\v{e}zd\'a\v{r}sk\'a ro\v{c}enka (Astronomer's Yearbook, further abbreviated as HR) has been published annually in the former Czechoslovakia (now Czech Republic) since 1921. The only exceptions were the years of World War II (1942--1945) when the series was interrupted. From the very beginning, HR was meant to serve mainly amateur astronomers. Therefore the ephemerides contained there were given with a limited precision --- arcseconds for the positions of the Sun and stars, arcminutes for the Moon and planets. A short look at the very first edition reveals that its contents were not much different from the present one: On 155 pages there were given positions and risings/settings of the Sun, Moon, and planets (Mercury -- Neptune), phases of the Moon, positions and phenomena of Galilean satellites of Jupiter and four satellites of Saturn, mean and apparent positions of bright stars, and also information on meteor showers. There were also given relatively detailed explanatory notes.

The first 20 editions (for the years 1921--1940) were prepared by B.~Ma\v{s}ek, vice-director of Ond\v{r}ejov Astronomical Observatory. All information dependent on geographical position was referred to the 15th meridian east and 50th parallel north. During these years, the contents of HR were gradually expanded ---  Pluto was added to the planets, and included were the predictions of solar and lunar eclipses and occultations, and information on variable stars and comets.  Since 1936 a new chapter, devoted to progress in astronomy, was introduced.

Beginning with HR for 1941, the preparation was taken over by V.~Guth and F.~Link (Astronomical Observatory Ond\v{r}ejov). Since the availability of foreign ephemerides became limited, some computations had to be made at home, but the contents did not change very much.  In 1952, J.~Bou\v{s}ka (Charles University Prague) joined, and gradually more and more co-authors became involved. Since 1956, J.~Bou\v{s}ka, V.~Guth and B.~Onderli\v{c}ka (Masaryk University Brno) became the leading authors. In 1960, the collective counted ten authors, and some important novelties were implemented: Ephemeris Time (instead of Universal Time used previously) was introduced as a time argument (see Section 1.2), and graphical representations of the positions of Jupiter's satellites were added. A detailed explanatory part was included, so that the book had 219 pages.

As the number of pages gradually grew (due to the growing information on the progress in astronomy), the annual publication of the explanatory part was removed in 1966.  J.~Ruprecht (Astronomical Institute Prague), responsible for the part on progress in astronomy, joined the team in 1967. In 1974, in response to user requirements, the explanatory notes were inserted again, which resulted in 279 pages.

During all these years, HR was heavily dependent on foreign ephemerides --- at the beginning, it was based on French ephemeris book {\it Connaissance des Temps}, later on the Soviet {\it Astronomitsheskii Jezhegodnik}, and eventually the Anglo-American {\it Astronomical Almanac} took over the role. Because this led sometimes to unpleasant delay in publishing HR, it was decided to compute the basic ephemerides at the Astronomical Institute, Czechoslovak Academy of Sciences. Thus, step by step between 1981 and 1986, all ephemerides of the Sun, Moon, planets and their satellites, and corresponding phenomena, started to be computed by J.~Vondr\'ak (Astronomical Institute Prague). To this end, his own subroutines, based on a re-calculated Improved Lunar Ephemeris, Bretagnon's analytical theory VSOP 1982 (later on, 1987), Sampson's theory (for the Galilean Jupiter's satellites) and elements given in the {\it Explanatory Supplement to the Astronomical Ephemeris} (for the four satellites of Saturn) were used. This practice has been maintained, with several minor improvements, until now.

Since 1993 HR is published as a co-edition by the Observatory and Planetarium Prague and Astronomical Institute of the Czech Academy of Sciences Ond\v{r}ejov.  The team of authors was coordinated by P.~P\v{r}\'{\i}hoda (Planetarium Prague, 1980--2010) and J.~Rozehnal (\v{S}tef\'anik Observatory Prague, since 2011). Significant changes were introduced, beginning with HR for 2011: the printed part was limited to only the most often used ephemerides (Sun, Moon, eclipses, planets, dwarf planets and asteroids, comets, meteors, variable stars, transiting exoplanets, occultations of stars brighter than 5th magnitude, and calendar of events, altogether some 120 pages), and much more detailed information is given on an attached CD.  The complete HR is now also available (for subscribers only) on the web at \mbox{http://rocenka.observatory.cz}.

\subsection{Russian Astronomical Yearbooks \, {\it (Marina Lukashova, Michael Sveshnikov )}}

The history of Russian astronomical ephemerides is closely connected
to development of the Russian fleet. The first ephemerides of solar
declinations for 1703-1729 were published by L.~F.~Magnitsky.
However, up to the beginning of the 20th century in Russia, there were only
small fascicle editions, as a rule, intended for nautical
astronavigation, which were arranged on the basis of foreign
publications.  Not until 1921 was the first Russian astronomical yearbook
issued, edited by B.~V.~Numerov.  Until 1960, the preparation of
Russian ephemerides for the {\it Astronomical Yearbook} (AY) was carried out in
cooperation with foreign scientists.  Since 1998, the preparation
has been carried out at the Institute for Applied Astronomy (IAA)
of the Russian Academy of Sciences.

The structure and information content of the
AYs have been modified several times.  The changes were caused both by
the development of techniques in astronomical observations and
achievements in fundamental astrometry and celestial mechanics. The last reform
of the AY was completed in 2008, according to the 2000--2006 resolutions of the IAU,
which essentially changed theoretical basis of the ephemeris
calculations. All relevant resolutions of the IAU have been implemented in the
ephemerides.  At present, the ephemerides in AY are referred to the classical
equinox system, but the parameters for conversion to the 
new system are also given.  Beginning in 2007, the domestic EPM theories 
(N-body numerical integrations computed by the IAA) are
accepted as the national standard for fundamental ephemerides by a
resolution of an all-Russian conference ``Coordinate, time and navigational
support'' (CTNS-2007).

Part of the data published in the AY are located on the Internet at \\
\mbox{http://www.ipa.nw.ru/PAGE/EDITION/RUS/rusnew.htm}.  In 2004, the
{\it Explanatory Supplement to the Astronomical Yearbook} was published (in
Russian) \citep{ESAY}.

In 1927, after a meeting of representatives of the naval departments in the
Astronomical Institute, it was resolved to create a special nautical
yearbook.  As a result, in 1930 {\it The Nautical Astronomical Yearbook}
(NAY) began to be issued.  Simultaneously with the development of aircraft,
{\it The Air Astronomical Yearbook} (AAY) began to be issued also (up to
1996).  For ships engaged in long-run sailing,  a new navigating manual
(biennial {\it The Nautical Astronomical Almanac}, NAA-2) was developed at IAA
and has been published since 2002. It is intended for solving the tasks of
nautical astronavigation, as well as NAY.  Comparison of the data in the AY and
NAY with their foreign analogues indicates their full accordance to a modern standard.

In 2007, an electronic version of AY was completed and named
``Personal Astronomical Yearbook''.  The PersAY system covers
the basic types of ephemerides published in the AY and also provides
the capability to calculate topocentric ephemerides.

An electronic version of NAA-2 has been developed for remote access, 
``The Shturman'', which is available at \mbox{http://shturman.ipa.nw.ru/maa}. It is
intended for solving basic astronavigation tasks included in the
NAA-2.  The solutions are carried out in accordance with the accepted accuracy
in these editions ($0.1'$) and formatted according to the protocols for such solutions,
as given by the published examples.

All ephemeris data are prepared on the basis of a multi-purpose
program complex for ephemeris calculations called ERA (see Section 1.4). Page layouts of tables
are formed with the help of system IZDATEL in the language of TeX. Both
systems were developed at IAA.

\subsection{Explanatory Supplement to the Astronomical Almanac \, {\it (S. E. Urban)}}

A primary goal of Commission 4 has always been to facilitate data exchanges.  Equally important to tabular data are the algorithms and primary data sources from which they are computed.  Throughout the 1930s, the British Nautical Almanac included detailed explanations, and by the end of that decade each volume was almost 1000 pages.  Although the explanations were of value to many astronomers, much of this information was unnecessary to the daily user of the publication, and there were complaints regarding the books being unwieldy.  Cutting the size of the yearly almanac by providing a more permanent explanatory supplement was under consideration at Her Majesty's Nautical Almanac Office (HMNAO) when World War II began.  At that time, ``a drastic cut was imposed on the overall size of subsequent editions by the exigencies of war" \citep{ES61}.
The preface to the 1942 almanac describes separating the ephemeral material from the permanent data and explanations, and also states, ``It is possible that publication of the Supplement will be delayed for some time."  In fact, it did not get published for 20 more years.

There was much work done on the Supplement shortly after the war.  However, the introduction of Ephemeris Time in 1950 (see Section 1.2), and a series of IAU resolutions in 1952 --- to  become effective in 1960 --- meant there was value in delaying the publication in order to describe the new material. Additionally, in 1954, the US Naval Observatory (USNO) and HMNAO took steps in unifying their respective nautical almanacs; thus, a ``delayed" supplement would be a joint publication of the two offices, with shared responsibilities.

The first edition, titled {\it Explanatory Supplement to the Astronomical Ephemeris and The American Ephemeris and Nautical Almanac}, was published in 1961 and reprinted with minor amendments in 1972, 1974, and 1977 \citep{ES61}.
Its primary function was ``to define precisely, for each individual ephemeris: the quantity tabulated; the fundamental data on which it is based; and how it is derived from those data" ({\it ibid.}).  It included many numerical examples.

Due to major changes, such as the IAU introduction of new astronomical constants (1976), new standard epoch and equinox (J2000.0), new time arguments (dynamical time\-scales), and new fundamental ephemerides (based on numerical integration), it was decided that a new supplement was needed. Work progressed throughout the 1980s with contributions from USNO, HMNAO, Jet Propulsion Laboratory (JPL), and Bureau des Longitudes. The second edition, titled, {\it Explanatory Supplement to the Astronomical Almanac} was published in 1992 \citep{ES92}.

Following 1992, several changes took place in positional astronomy. Advances in radio observations allowed the celestial reference frame to be tied to extragalactic radio sources instead of bright nearby stars, thus the ICRS replaced the FK5 system.  The success of the Hipparcos satellite dramatically altered observational astrometry.  Improvements in Earth orientation observations led to new precession and nutation theories.  Additionally, a new positional paradigm, no longer tied to the ecliptic and equinox, was accepted.  These changes culminated in a series of resolutions passed at IAU GAs between 1997 and 2006 (see Section 1.1).  Largely because of these resolutions, the staffs at USNO and HMNAO decided a new supplement was needed.  The third edition, also titled {\it Explanatory Supplement to the Astronomical Almanac}, was published in 2012 \citep{ES12}.
It contains contributions from USNO, HMNAO, JPL, Observatoire de Paris, US Geological Survey, National Geospatial-Intelligence Agency, Lohrmann Observatory, and a handful of universities.

\subsection{Impact of the 1997-2006 IAU Resolutions on the National Almanacs\\{\it (J.~A.~Bangert)}}

Resolutions adopted by the XXIIIrd and XXIVth IAU GAs (Kyoto, 1997, and Manchester, 2000) were perhaps the most significant set of international agreements in positional astronomy in several decades and arguably since the Paris conference of 1896.  These resolutions and additional resolutions passed in 2006 had a profound impact on the production, design, and content of the national almanacs. The adoption of the International Celestial Reference System (ICRS), new barycentric and geocentric celestial reference systems, new solar system ephemerides aligned to the ICRS, new consistent models for precession and nutation, new reference points on the Earth and sky for measuring Earth rotation, new times scales, and new nomenclature all required major revisions of the data sets and software used to produce the almanacs, and necessitated radical changes to the layout and content of many data tables. At a deeper level, the resolutions introduced changes to long-standing concepts, such as the definition of right ascension, which required additional explanatory material in order to make the changes understandable to the user. To add to the challenge, the 2000 resolutions on precession-nutation, the celestial intermediate pole, and the celestial and terrestrial intermediate origins called for their implementation on 1~January 2003, less than three years after their passage (in the original resolutions, the names of these points were different).

At the time they were passed, the 2000 IAU resolutions were not ready for implementation. It can be argued that it took six more years before they were. For example:

\begin{itemize}
\item Resolution B1.6 in 2000 adopted a new precession-nutation model based on work that would not be peer-reviewed and published until 2002 \citep{mathews02}. The actual nutation series, necessary for implementing the model, was not a part of this paper. Software implementations of the model became available in early 2001, and were included in the IERS Conventions (2003) \citep{iers03} and in release 2 (in April 2003) of the IAU's Standards of Fundamental Astronomy (SOFA) software library.

\item  Resolution B1.6 in 2000 encouraged development of ``new expressions of precession consistent with the IAU 2000A nutation model." Such a precession model, along with a new definition of the ecliptic, was adopted by Resolution~1 passed by the XXVIth IAU GA (Prague, 2006). An interim precession model was published in the IERS Conventions (2003) and appeared in release~2 of SOFA, but was never adopted by an IAU resolution.

\item The values in the ``frame bias" matrix, which provides the critical tie between the mean dynamical equator and equinox at J2000.0 and the ICRS, were published in the IERS Conventions (2003).

\item The radical changes adopted in 2000 ultimately required new nomenclature, which became part of Resolution 2, ``Supplement to the IAU 2000 Resolutions on Reference Systems," passed by the GA in Prague in 2006.  Some of the nomenclature introduced in the 2000 resolutions was changed in the 2006 resolution.
\end{itemize}

The complex nature of the 2000 IAU resolutions, their evolving realization after the time of passage, their impact on almanac content, and the deadlines involved in almanac production made it very difficult if not impossible to meet the IAU's specified date for implementation.   The plans of many of the national almanac offices were described at the IAU Division I meeting at the XXVth IAU GA (Sydney, 2003). It was clear that the resolutions had to be implemented carefully and gradually over time. 
An additional complication for \textit{The Astronomical Almanac} --- a joint publication of the almanac offices of the United Kingdom and the United States --- was an action taken by the Council of the American Astronomical Society in 2004 expressing ``the concerns of the US astronomical community about the implementation of the new origins and coordinates adopted by the IAU by resolutions in 1997 and 2000." Among other provisions, the action asked ``the US Naval Observatory to continue to provide information in the prevailing coordinate system in the Astronomical Almanac."

The table below gives the major milestones in the timeline over which the IAU resolutions of 1997--2006 were implemented in \textit{The Astronomical Almanac}.

\begin{center}
\begin{tabular}{p{0.15\linewidth}p{0.8\linewidth}}\hline
Edition Year & \multicolumn{1}{c}{Milestone}\\
\hline
\multicolumn{1}{c}{1999} & Added table of ICRF radio-source positions\\
\multicolumn{1}{c}{2003} & Implemented JPL DE405/LE405, aligned to ICRS, as solar system ephemeris\\
\multicolumn{1}{c}{2006} & Implemented IAU resolutions of 2000 (IAU 2000A precession-nutation from IERS (2003))\\
\multicolumn{1}{c}{2009} & Implemented IAU resolutions of 2006\\
\hline
\end{tabular}
\end{center}
\NOPRINT{
\begin{itemize}
\item[] 1999 \,\, Added table of ICRF radio-source positions
\item[] 2003 \,\, Implemented JPL DE405/LE450, aligned to ICRF, as solar system ephemeris
\item[] 2006 \,\, Implemented IAU resolutions of 2000 (IAU 2000A precession-nutation from IERS Conventions (2003))
\item[] 2009 \,\, Implemented IAU resolutions of 2006
\end{itemize}
}

\section*{Summary \, {\normalsize\it (C. Hohenkerk)}}

This summary of the work of Commission 4 over the last century reflects the interrelationship between reference systems, time scales, astronomical constants, observations, and the various methods used for computing ephemerides, including  the modern integrations.  As well, it highlights the important relationship between the ephemerides themselves and the publications, i.e., the national ephemerides and almanacs, which particularly in the early days, before the age of personal computers, were the main media for dissemination. However, importantly it demonstrates the benefit from the collaboration between the Commission~4 members, their institutes and other groups both the inside and outside the IAU. 
 
This cooperation of experts combined with the requirements of all users of ephemerides is well demonstrated by the last Commission~4 Working Group on Standardizing Access to Ephemerides and File Format Specification, which was chaired by James Hilton.  Their final report \citep{hilton15} was presented at the recent GA, and the detailed specifications are at arXiv.org  \mbox{(http://arxiv.org/abs/1507.04291)}.

I would like to thank all the authors who collaborated in drafting the various sections of these ``Historical Reflections'' as well as those who reviewed them, who provided many helpful suggestions.  Thanks are also due to our Organising Committee. Finally, I would like to thank our editor, Past President George Kaplan. George has not only written some sections, he has also put considerable effort into making this report very interesting and worthwhile. We were very fortunate that we had someone with his background knowledge who was willing to put this together, and we all thank him for his efforts.

\bigskip
\bibliography{Com4_Historical_Reflections}

\end{document}